\documentclass[11pt]{article}
\pdfoutput=1

\usepackage[T1]{fontenc} 
\usepackage[utf8]{inputenc}
\usepackage{graphicx,multirow,subfigure}
\usepackage{float}
\usepackage{bm}
\usepackage{hyperref}
\usepackage{amsmath}
\usepackage{amssymb}
\usepackage{amscd}
\usepackage{latexsym}
\usepackage{slashed}
\usepackage{color}
\usepackage{graphicx}
\usepackage[normalem]{ulem}
\usepackage{color}
\definecolor{orange}{cmyk}{0,0.5,1,0}
\usepackage{verbatim}
\usepackage{rotating}

\newcommand{\nn}{\tilde{\nu}}
\def\lsim{\raise0.3ex\hbox{$\;<$\kern-0.75em\raise-1.1ex\hbox{$\sim\;$}}}
\def\gsim{\raise0.3ex\hbox{$\;>$\kern-0.75em\raise-1.1ex\hbox{$\sim\;$}}}

\usepackage{cancel}
\usepackage[utf8]{inputenc}

\def\be{\begin{equation}}
\def\ee{\end{equation}}
\def\bea{\begin{eqnarray}}
\def\eea{\end{eqnarray}}
\def\nn{\nonumber}
\def\Be{{}^8\textrm{Be}}

\begin{document} 
\title{\bf New Physics Suggested by Atomki Anomaly}

\author{\\[-5mm]
Luigi Delle Rose$^{a,b}$, Shaaban Khalil$^{c}$, Simon J. D. King$^{b,d}$, Stefano Moretti$^{b,e}$ \\[0.35cm] 
\small {\it $^a$INFN, Sezione di Firenze, and Department of Physics and Astronomy, University of Florence,} \\
\small{\it Via G. Sansone 1, 50019 Sesto Fiorentino, Italy}\\
\small{\it $^b$School of Physics and Astronomy, University of Southampton,} \\
\small{\it Highfield, Southampton SO17 1BJ, United Kingdom}\\
\small{\it $^c$Center for Fundamental Physics, Zewail City of Science and Technology, 6 October City, Giza 12588, Egypt}\\
\small{\it $^d$INFN, Sezione di Padova, and Dipartimento di Fisica ed Astronomia G. Galilei, Universit\`a di Padova,}\\ 
\small{\it Via Marzolo 8, 35131 Padova, Italy}\\
\small{\it $^e$Particle Physics Department, Rutherford Appleton Laboratory,} \\
\small{\it Chilton, Didcot, Oxon OX11 0QX, United Kingdom}\\
}

\maketitle

\begin{abstract}
	We consider several extensions of the Standard Model (SM) which can explain the anomalies observed by the Atomki collaboration in the decay of  excited states of Beryllium via a new boson with a mass  around 17 MeV yielding $e^+e^-$ pairs. We show how both  spin-0 and 1 solutions are possible and describe the  Beyond the SM (BSM) scenarios that can accommodate these. They include BSM frameworks with either an enlarged Higgs, or gauge sector, or both.
\end{abstract}

\flushbottom

\section{Introduction}
\label{sec1:intro}

{{The quest for New Physics (NP) above and Beyond the Standard Model (BSM)  has always seen a twofold approach. On the one hand, the high energy frontier has been pursued, typically through multi-purpose experiments at hadron accelerators, like the $Sp\bar pS$, Tevatron and LHC. On the other hand, the high precision frontier has been exploited, typically at lepton collider experiments, like LEP and SLC. Alongside this time honoured two-prong pursuit, over the years, a transversal dimension, covering both hadron and lepton colliders, centered on flavour physics, has also developed. So that, presently, the attention of the particle physics community in unveiling some NP has mainly been concentrated  upon these three research strands. However, surprises may arise in other contexts, notably from (much) lower energy experiments. In this respect, results from $(g-2)$ of the muon are prototypical. Another   interesting result which has recently been reported is the one in Ref. 
 \cite{Krasznahorkay:2015iga} {{(see also \cite{Krasznahorkay:2017gwn,Krasznahorkay:2017bwh,Krasznahorkay:2017qfd,Krasznahorkay:2018snd})}}, by the Atomki experiment  \cite{Gulyas:2015mia}. 
The latter is a pair spectrometer for measuring multi-polarities of  nuclear transitions, specifically, using a multi-detector array designed and constructed for the simultaneous measurement of energy and angular correlations of electron-positron pairs, in turn emerging via  internal pair creation from a variety of nuclear transitions in various isotopes, such as $^{16}$O, $^{12}$C and $^8$Be.
The intriguing result reported in  \cite{Krasznahorkay:2015iga} concerns $e^+e^-$ correlations  measured for the isovector magnetic dipole 17.64 MeV state (with spin-parity and isospin, $J^P=1^+$, $T=1$, respectively),
and the isoscalar magnetic dipole 18.15 MeV state ($J^P =1^+$, $T=0$) in their transitions to the ground state ($J^P =0^+$, $T=0$)  for the Beryllium case. Significant deviations from the internal pair creation rate were observed at large angles in the angular correlation for the isoscalar transition with a confidence level of more than $5\sigma$. This observation may  indicate that, in an intermediate step, a (light) neutral boson with a mass of $16.70\pm0.35\,({\rm stat})\pm0.5\,({\rm sys})$  MeV has been created. In fact, also the 17.64 MeV transition eventually appeared to present a similar anomaly, albeit less significant, with a boson mass broadly compatible with the above one, i.e., $17.0\pm 0.5\, ({\rm stat})\pm 0.5\, ({\rm sys})$ MeV\footnote{It should however be mentioned that this second anomaly was never documented in a published paper, only in proceedings contributions.}. }}

{{The purpose of this review is to discuss possible solutions to these results, assuming that the neutral boson could be either a spin-1 or spin-0 object, belonging to a variety of BSM scenarios. The plan is as follows. In the next section we consider the characteristics of the results reported by the Atomki experiment. Then we describe possible candidate particles for such a light bosonic state. Finally, we illustrate the embedding of such solutions in possible theoretical models, in presence of a variety of experimental constraints emerging from both low and high energy experiments. We finally conclude.}}

\section{The Atomki experiment and 17 MeV Beryllium anomaly}

The Atomki pair spectrometer experiment \cite{Gulyas:2015mia} was set up for searching $e^+ e^-$ internal pair
creation in the decay of excited $^8$Be nuclei (henceforth, $^8{{\rm Be}^*}$), the latter being produced with the help of a beam
of protons directed on a $^7$Li target. The proton beam was tuned in such a way that the
different $^8$Be excitations could be separated in energy with high accuracy.

\begin{figure}[t]
\centering
\includegraphics[scale=0.5]{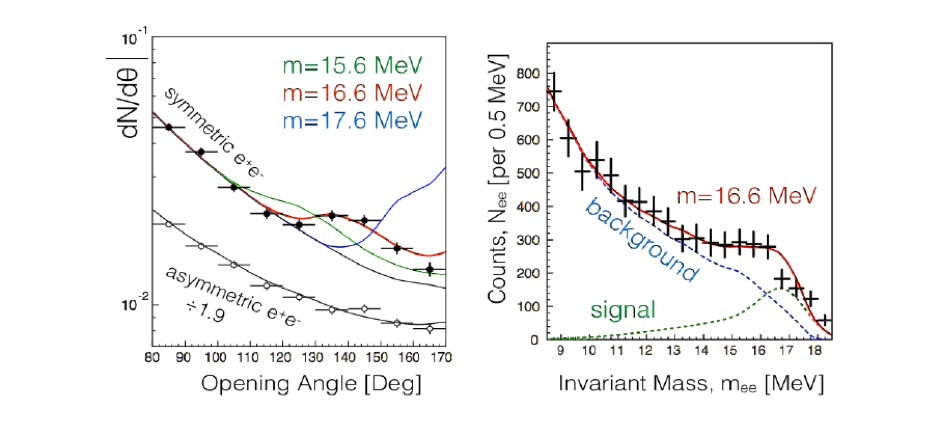}
\vskip -0.75cm
\caption{Angular and invariant mass distributions of the internal conversion
electron-positron pairs measured by the Atomki spectrometer (from \cite{Feng:2016ysn}).}
\label{invmass}
\end{figure}

In the data collection stage, a clear anomaly was observed in the decay of $^8{{\rm Be}^*}$ with
 $J^P=1^+$ into the ground state $^8{\rm Be}$ with spin-parity $0^+$ (both with $T=0$),
where $^8{{\rm Be}^*}$ had  an excitation energy of 18.15~MeV \cite{Krasznahorkay:2015iga}. Upon analysis of the electron-positron properties, 
the spectra of both their opening
angle $\theta$ and  invariant mass $M$ presented the characteristics of  an excess consistent with an intermediate boson (henceforth, $X$)
being produced on-shell
 in the decay of  the $^8{{\rm Be}^*}$ state, with the $X$ object subsequently decaying into $e^+e^-$ pairs.
As mentioned, the best fit to the mass $M_X$ of $X$ was given as 
$
M_X = 16.7 \pm 0.35\ \text{(stat)}\ \pm 0.5\ \text{(sys)\ MeV},
$
 \cite{Krasznahorkay:2015iga}
in correspondence of a ratio of Branching Ratios (BRs) obtained as 
\be
{\mathcal B}\equiv \frac{{\rm BR}(^8{{\rm Be}^*} \to X + {^8{\rm Be}})}{{\rm BR}(^8{{\rm Be}^*} \to \gamma + {^8{\rm Be}})} \times {\rm BR}(X\to e^+ e^-) 
= 5.8 \times 10^{-6}.
\label{eq:BeAnomaly}
\ee
 The signal appeared as a bump over the monotonically decreasing background from pure Quantum Electro-Dynamics (QED) interactions, i.e.,  internal pair creation via $\gamma^*\to e^+e^-$ splittings. This excess appeared only for symmetric energies of $e^+ e^-$, 
 as expected from an on-shell non-relativistic particle. In addition, the opening angle of electron-positron pair and their invariant mass distributions presented the characteristics of an excess consistent with an intermediate boson. The measurements yielded  the mentioned value $M_X$ from the  invariant mass $m_{e^+ e^-}$,
in correspondence of an angular excess around  $\sim 135^\circ$,  as shown in Fig.~\ref{invmass}.
The best fit to data was obtained for a new particle interpretation, in which case the statistical significance of the excess is 6.8
sigma. {{The aforementioned result from the 17.64 MeV transition yielded $M_X=17.0\pm 0.5\, ({\rm stat})\pm 0.5\, ({\rm sys})$ as best fit, in correspondence of an angular peak around 
$155^\circ$ with ${\mathcal B}=4.0\times10^{-6}$. The corresponding significance is nowhere near discovery though.}}

\section{Candidates for the new boson}

An explanation of the nature of the intermediate particle, $X$, decaying to electron-positron pairs, was attempted by considering 
it as boson either with spin zero (scalar or pseudoscalar) or with spin one (vector or axial-vector). We introduce all possible combinations in turn.

\subsection{Scalar particle}

If the intermediate particle $X$ is a scalar, $\phi$ ($J^P = 0^+$), then the decay $^8{\rm Be}^*(1^+) \to {}^8{\rm Be}(0^+) + \phi$ implies, due to angular momentum conservation, that $\phi$ should have $L=1$. Also, from parity conservation, it must have a parity equal to $(-1)^L$, which is $-1$ and this contradicts the assumption that $\phi$ is scalar with even parity. Therefore, one can conclude that a scalar intermediate particle is ruled out.

\subsection{Pseudoscalar particle}

{{The situation is different if the intermediate particle is a pseudoscalar, $A$ ($J^P = 0^-$) \cite{Ellwanger:2016wfe}.  In this case, given the quantum numbers of the $^8$Be$^*$ and $^8$Be
states, the intermediate boson can indeed be a $J^P = 0^-$ pseudoscalar particle if it was emitted with $L = 1$ orbital momentum. It was in fact shown in Ref. \cite{Ellwanger:2016wfe} that $A$ can account for the Atomki results if its Yukawa couplings with the SM fermions are of order of the Yukawa couplings of the SM Higgs.}}

\subsection{Vector particle}

A neutral vector boson is the most common example considered for explaining this signal \cite{Feng:2016jff,Feng:2016ysn,Gu:2016ege,Chen:2016dhm,Liang:2016ffe,Jia:2016uxs,Kitahara:2016zyb,Chen:2016tdz,Seto:2016pks,
Neves:2016ugb,Chiang:2016cyf}. It was emphasised that it can be a valid candidate if its 
coupling is constrained as $g' \sim 10^{-3}$. 

\subsection{Axial-Vector particle}

The pure axial-vector boson is also considered and it was shown that it can be a candidate if its coupling satisfies $g' \sim 10^{-4}$, as done in \cite{Kozaczuk:2016nma,Feng:2016ysn,Kahn:2016vjr}. The case of general spin-one boson, with no definite parity, {i.e.}, it is a mix of vector and axial-vector, could be a possible candidate after taking care of stringent constraints from atomic parity violation.

The couplings of these new light bosons with the SM particles remain an open question and subject 
to severe constraints from several experiments. 

\section{Experimental constraints on the  pseudoscalar explanation}


The reduced couplings $\xi_q$ of a pseudoscalar $A$ to quarks is defined as 
\begin{equation}
{\mathcal L}_{Aqq} = \xi_q \frac{m_q}{v} A \bar{q} i \gamma_5 q,
\end{equation}
with $v\sim 246$~GeV.  {{Assuming such fundamental interactions and adopting the  nuclear
shell model wave functions with definite isospin $T = 0$ of Ref.~\cite{Ellwanger:2016wfe}, one finds that 
\begin{equation}\
\label{xiud}
\xi_u + \xi_d \approx  0.6
\end{equation}
or, for $\xi_u=\xi_d\equiv \xi$, $\xi \approx  0.3$. Furthemore, if $A$ has Yukawa couplings to quarks and leptons which are proportional to
the Yukawa couplings of the SM Higgs boson rescaled by generation
independent factors $\xi_d \approx \xi_u \approx \xi_e$ (or $\xi_u \ll \xi_d$),
and the Yukawa couplings to BSM fermions are not much larger than the electric
charge $e$, $A$ has a BR of
about 99\% into $e^+ e^-$ and only about 1\%
into $\gamma\gamma$. Its total width is then dominated by $A\to e^+ e^-$
and given by
\begin{equation}
\Gamma(A) = \xi_e^2 \frac{m_e^2}{8\pi v^2}M_A = \xi_e^2\cdot 2.9\times 10^{-15}\ \text{GeV}
\end{equation}
for $M_A=17$~MeV. Its decay length is
\begin{equation}
l_A=\frac{p_A}{M_A \Gamma(A)}\; .
\end{equation}
For the decay $^8{{\rm Be}^*}\to{^8{\rm Be}}+A$ with $M(^8{{\rm Be}^*})-M(^8{\rm Be})=18.15$~MeV
we obtain
\begin{equation}
l_A \sim \frac{1}{\xi_e^2}\cdot  2.5\ \text{cm}.
\end{equation}
(For $M_A=17.9$~MeV, $2\,\sigma$ above the central value in $M_X$ from the 18.15 MeV transition,
we obtain $l_A \sim \frac{1}{\xi_e^2}\times 1.1\ \text{cm}$.)
In order to explain the observed anomaly in the Atomki pair spectrometer
experiment \cite{Krasznahorkay:2015iga}, $l_A$ should then not be much larger
than 1~cm leading to
\begin{equation}
\label{xiegt1}
\xi_e {\gsim} 1\; ,
\end{equation}
depending somewhat on the precise value of $M_A$.
}}

Light pseudoscalars are subject to constraints from searches for axions or
axion-like particles. For recent summaries of constraints relevant
for light pseudoscalars decaying dominantly into $e^+ e^-$, see
\cite{Dolan:2014ska,Andreas:2010ms,Essig:2010gu,Hewett:2012ns,Dobrich:2015jyk}. 
However, since we allow for different Yukawa type couplings rescaled by $\xi_u$,
$\xi_d$ and $\xi_e$ with respect to SM Higgs couplings, at least some experimental
constraints studied therein have to be reconsidered. Constraints
from $\pi^0\to \gamma + X$ from the NA48/2 experiment, which play a major role
for the $Z'$ scenario \cite{Feng:2016jff,Feng:2016ysn}, do not apply here
since the decay $\pi^0\to \gamma + A$ would violate parity. 

Constraints  also originate from
flavour violating meson decays, analysed recently in \cite{Dolan:2014ska}, and are mainly due to 
the following decays: $K^+ \to \pi^+ + A$ (constrained by the $K_{\mu 2}$ experiment
\cite{Yamazaki:1984vg}),
$K^+ \to \pi^+ + {\sl invisible}$ (measured by the experiments E787 \cite{Adler:2004hp} and
BNL-E949 \cite{Artamonov:2009sz}),
$B_s\to \mu^+\mu^-$ (measured by the LHCb collaboration \cite{Aaij:2013aka}
and the CMS collaboration \cite{Chatrchyan:2013bka}, see \cite{CMS:2014xfa}
for a LHCb/CMS combination) and
$B^0\to K^0_S + invisible$ (measured by CLEO \cite{Ammar:2001gi}). It turns out that the most stringent Flavour Changing Neutral Current (FCNC)  constraint is due to 
 $K^+ \to \pi^+ + A$ from the $K_{\mu 2}$ experiment \cite{Yamazaki:1984vg}. This process depends on a
loop-induced $Asd$ vertex (with $W$ bosons and up-type quarks in the
loop) which depends, in turn, on the couplings of $A$
to $d$- and $u$-type quarks. 
Constraints from \cite{Yamazaki:1984vg} can lead to 
\begin{equation}
\label{boundxid}
\xi_d \lsim 2\times 10^{-2}.
\end{equation}
A similar constraint can be obtained from the process $B \to K +A$.

Constraints from searches for $K^+ \to \pi^+ + {\sl invisible}$ from E787 and BNL-E949
\cite{Adler:2004hp,Artamonov:2009sz} apply only if $A$ decays outside the detectors,
i.e., if $\xi_e$ is small enough. According to \cite{Andreas:2010ms}, identifying
now $C_{Aff}$ in \cite{Andreas:2010ms} with $\xi_e$, this is not the case for $\xi_e \gsim 0.3$.

According to \cite{Dolan:2014ska}, the constraints from $B_s\to \mu^+\mu^-$
(through an off-shell $A$) rule out any $\xi \gsim 0.7$ which is
weaker than the constraint \eqref{boundxid} from $K^+ \to \pi^+ + A$.
Again, the loop contributions to the $Asb$ vertex considered in \cite{Dolan:2014ska}
are incomplete within an Ultra-Violet (UV) complete extension of the Higgs sector, and
could again be cancelled by additional BSM contributions as
in the case of the $Asd$ vertex.

The constraints from $B^0\to K^0_S + invisible$ measured by CLEO \cite{Ammar:2001gi}
apply only if the pseudoscalar $A$ produced in $B^0\to K^0_S + A$ decays outside
the detector. Accordingly these constraints depend both on the BR$(B^0\to K^0_S + A)$,
hence on the $Asb$ vertex or on $\xi_u,\xi_d$, and on the $A$ decay length which depends on
$\xi_e$. These quantities are identified in \cite{Dolan:2014ska} where a limit $\xi \gsim 3.5$ on all flavours satisfies the constraints, since then the $A$ decay
length becomes short enough despite the large production rate. Using this constraint
only for $\xi_e$ is conservative, if $\xi_u,\xi_d < \xi_e$ is assumed.

Finally, $\xi_e \gsim 3.5$ satisfies also bounds on $A$ production in radiative $\Upsilon$ decays
$\Upsilon \to \gamma + invisible$ interpreted as $\Upsilon \to \gamma + A$
from CLEO \cite{Balest:1994ch} and BaBar \cite{Aubert:2008as}, which apply only if
$A$ decays outside the detectors. For $M_A\sim 17$~MeV, following \cite{Andreas:2010ms},
this is not the case for $\xi_e \gsim 1.5$.

Other important constraints on light pseudoscalars originate from beam dump
experiments. From the Orsay experiment of Ref. \cite{Davier:1989wz},   lifetimes $\tau_A$
in the range $5\times 10^{-12}\ \text{s} \lsim \tau_A \lsim 2\times 10^{-9}$~s
are ruled out for $M_A\sim 17-18$~MeV. This has already been translated into constraints
on a reduced pseudoscalar-fermion Yukawa coupling $C_{Aff}$ in \cite{Andreas:2010ms},
where $C_{Aff} = \xi_e$ in our notation. Following \cite{Andreas:2010ms},
$0.4 \lsim C_{Aff} \lsim 4$ is ruled out by this constraint.
Since $\xi_e < 0.4$ is incompatible with \eqref{xiegt1}, one is left with
\begin{equation}
\label{orsay}
 \xi_e {\gsim} 4\; .
\end{equation}
This constraint leads automatically to the satisfaction of the lower bound
$\xi_e \gsim 3.5$ from $B^0\to K^0_S +\ invisible$, as well as to a short enough
decay length \eqref{xiegt1} for the Atomki pair spectrometer experiment. 
It is also compatible with the exclusion from the NA64 experiment \cite{Banerjee:2018vgk} provided that $\xi_e \lesssim 15$.

Another potentially relevant experiment is the proton beam dump on copper CHARM
experiment \cite{Bergsma:1985qz}. In \cite{Bergsma:1985qz} constraints were
derived assuming that the production cross section and decay length of light
pseudoscalars correspond to those of axions, which is not the case here.
Relevant is the analysis in \cite{Dolan:2014ska} which uses the production
of light pseudoscalars in $K\to \pi + A$ and $B \to X + A$ decays. For
universally rescaled Yukawa couplings the region $\xi \gsim 1$
satisfies the constraints, since then the decay length of $A$ is too short to
reach the decay region of the CHARM experiment. This
constraint does not supersede the one in eq. \eqref{orsay}.

\section{Explanation of the Beryllium anomaly with a Pseudoscalar}

One of the less well studied solutions is that of the pseudoscalar, but this has been done in \cite{Ellwanger:2016wfe}. It was initially dismissed by \cite{Feng:2016jff,Feng:2016ysn} and subsequent authors by the argument that for such axion-like pseudoscalars $A$, fermion loops generate couplings of the form $g_{A\gamma\gamma} A F^{\mu\nu}(\gamma)\tilde{F}_{\mu\nu}(\gamma)$
which are strongly constrained by axion searches. However, light pseudoscalars
in this mass range
with tree level Yukawa couplings to electrons decay dominantly into electron-positron
pairs, unless Yukawa couplings to other charged fermions $f$ with mass $m_f$ are much larger
than $m_f/m_e$ compensating $g_{A\gamma\gamma}\approx 1/(8\pi m_f)$. For solutions to the Atomki anomaly, we require such couplings to electrons and hence one should dismiss the pseudoscalar solution.

To summarise the previous section investigating the constraints, couplings of the form $\xi _u + \xi _d \sim 0.6$ and $\xi_e > 4$ should satisfy all aforementioned constraints and provide an explanation to the Atomki anomaly, with the caveat that FCNCs must be suppressed by loop contributions at the level of at least $10\%$. 

{{Ultimately, it will be the Atomki experiment itself which will be in a position to either confirm or disprove the light pseudoscalar hypothesis. In fact,  the experiment is
currently  planning to study the $\gamma\gamma$ decays of the 17 MeV particle, also in $4He\to\gamma\gamma$ \cite{Krasznahorkay:2017bwh}, in order
to distinguish between a vector boson and  pseudoscalar boson scenario. According to the
Landau-Yang theorem, the (on-shell) decay of a vector boson by double $\gamma$-emission is forbidden, however, the
decay of a pseudoscalar one is allowed \cite{Moretti:2014rka}. The angular correlation of the $\gamma$-rays will be measured by
using 15 large ($3"\times3"$) LaBr$_3$ detectors. If the $A$ boson with a mass of 17 MeV is created in the
decay of the $J^P=0^-$ state and in turn decays into two $\gamma$-rays, their angular correlation $\theta$ should peak at
\begin{equation}
\cos\theta = 1-\frac{M^2_A}{2E_{\gamma}E_{\gamma^\prime}},
\end{equation}
where $M_A$ is the mass of the $A$ boson (17 MeV) and $E_{\gamma,\gamma^\prime}$ are the energies of the two photons. However, it should be kept in mind that a light pseudoscalar with tree level coupling to electrons would have a loop-induced BR to di-photons of only one percent or so, hence hardly visible with current Atomki data sets. At any rate, 
results in this respect, are eagerly awaited.}}


\section{Experimental constraints on the spin-1 explanation }
\label{sec:constraints}

Let us assume that the generic coupling of a new vector boson, $Z'$, to the SM fermions is given by the following interaction Lagrangian
\be 
-{\mathcal L_{\textrm{int}}} = Z'_\mu \sum_f \bar{\psi}_f  \gamma^\mu \left(C_{f,V}+ \gamma_5 C_{f,A}\right) \psi_f \,.
\ee

\subsection*{Experimental constraints on the lepton couplings}

We have not seen a $Z'$ in the electron beam dump experiment SLAC E141. Therefore, a $Z'$ has not been produced, hence
\be
C_{e,V}^2 + C_{e,A}^2 < 10^{-17}
\ee
or, else, a $Z'$ has been caught in the dump, hence
\be 
\frac{C_{e,V}^2 + C_{e,A}^2}{\textrm{BR}(Z' \to e^+ e^-)} \gsim  3.7 \times 10^{-9}. 
\ee 
We have not seen a $Z'$ either in the electron beam dump experiment NA64 \cite{Banerjee:2018vgk}. If a $Z'$ has been caught in the dump, this places the (stronger than E141) condition
\be
\frac{C_{e,V}^2 + C_{e,A}^2}{\textrm{BR}(Z' \to e^+ e^-)} \gsim 1.6 \times 10^{-8} \,.
\label{eq:NA64}
\ee

The parity-violating M{\o}ller scattering measured at the SLAC E158 experiment \cite{Anthony:2005pm} imposes a constraint on the product $C_{e,V} C_{e,A}$ of the $Z'$, namely 
\be 
\vert C_{e,V}  C_{e,A} \vert \lsim 10^{-8},
\ee
for $M_{Z'} \simeq 17$ MeV \cite{Kahn:2016vjr}. 

Furthermore, there could be contributions of a $Z'$ to the magnetic moments of electron and muon. The one-loop ones $\delta a_{l}$, mediated by a $Z'$, lead to 
\be
\delta a_{l}= \frac{r_{m_l}}{4\pi^2} \left[ C_{l,V}^2  \, g_V(r_{m_l}) - C_{l,A}^2  \, g_A(r_{m_l}) \right],
\ee
where $r_{m_l} \equiv (m_l/M_{Z'})^2$ and $g_V,g_A$ are given by%
\begin{align}
 g_V(r)= \int^1_0 dz\frac{z^2(1-z)}{1-z+r z^2}, \qquad
 g_A(r)= \int^1_0 dz\frac{(z-z^2)(4-z) + 2r z^3}{1-z+r z^2} \,.
\end{align}
The light boson contribution to the anomalous magnetic moment of the electron is required to be within the $2\sigma$ uncertainty of the departure of the SM prediction from the experimental result \cite{Giudice:2012ms}. Concerning the muon
anomalous magnetic moment \cite{Altmannshofer:2016brv}, which has been measured at
 Brookhaven National Laboratory (BNL) to a precision of $0.54$ parts per million, the current
average of the experimental results is given by \cite{Bennett:2006fi,Blum:2013xva,Lindner:2016bgg} 
\be
a^{\rm exp}_{\mu}=11 659 208.9(6.3)\times 10^{-10},
\ee
which is different from the SM prediction by $3.3\sigma$ to $3.6\sigma$: 
$
\Delta a_{\mu}=a^{\rm exp}_{\mu}-a^{\rm SM}_{\mu}=(28.3 \pm
8.7\ {\rm to}\ 28.7 \pm
8.0)\times 10^{-10}.
$
 We require again that the contribution of a $Z'$ to  $(g-2)_\mu$, which is mainly due to its axial-vector component, is less than the $2\sigma$ uncertainty of the discrepancy between the SM result and the experimental measure. 
For $M_{Z'} \simeq 17$ MeV, one then finds 
\bea 
\delta a_{e} &=& 7.6 \times 10^{-6} C_{e,V}^2  -3.8 \times 10^{-5} C_{e,A}^2 \simeq -10.5(8.1) \times 10^{-13},\\
\delta a_{\mu} &=&0.009 C_{\mu,V}^2 -C_{\mu,A}^2 \leq  2.9 (90) \times 10^{-9}.
\eea

Electron-positron colliders (like KLOE2) would be sensitive to a new spin-1 gauge boson via the channel $e^+ e^- \to \gamma, Z, Z' \to e^+ e^-$. From this process one finds 
\be 
(C_{e,V}^2 + C_{e,A}^2) {\rm BR}(Z' \to e^+ e^-) \lsim 3.7 \times 10^{-7}.
\ee

Similarly, $Z'$ contributions to neutrino-electron scattering implies a bound on the product of the electron and neutrino couplings to the $Z'$ \cite{Deniz:2009mu, Bilmis:2015lja}.

\subsection*{Experimental constraints on the quark couplings}

The couplings of  a light $Z'$ state with quarks are, in general, strongly constrained from $\pi^0 \to Z' +\gamma$
searches at the NA48/2 experiment \cite{Raggi:2015noa}. The process is proportional to the anomaly factor $ N_\pi = \frac{1}{2} (2 C_{u,V} + C_{d,V})^2$.
Therefore, one gets the following limit:
\be
\vert 2 C_{u,V} + C_{d,V} \vert \lsim \frac{3.6 \times 10^{-4}}{\sqrt{\textrm{BR}(Z' \to e^+ e^-)}}
\ee
for $M_{Z'} \simeq 17$ MeV. The contribution of the axial components is induced by chiral symmetry breaking effects and is, therefore, suppressed by the light quark masses.

Furthermore, atomic parity violation in Cesium (Cs) must be considered. In fact, very strong constraints on a light $Z'$ can be extracted from the measurement
of the effective weak charge of the Cs atom \cite{Davoudiasl:2012ag,Bouchiat:2004sp}:
\be 
\Delta Q_w = \frac{-2\sqrt{2}}{G_F} C_{e,A} \left[C_{u,V} (2 Z + N) + C_{d,V} (Z + 2 N) \right] \left( \frac{0.8}{(17~ {\rm MeV})^2}\right) \lsim 0.71
\ee
at $2\sigma$ \cite{Porsev:2009pr}.

\section{A $U(1)^\prime$ extension of the SM with a light and weakly interacting $Z'$}

We consider a generic  extension to the SM described by a new Abelian group $U(1)'$ \cite{Fayet:1980rr,Fayet:1980ad,Fayet:1990wx,Fayet:2007ua,Fayet:2008cn,Fayet:2016nyc,DelleRose:2017xil}. 
Due to the presence of two  such Abelian symmetries, $U(1)_{Y} \times U(1)'$, 
the most general kinetic Lagrangian of the corresponding fields, $\hat B_\mu$ and $\hat B'_\mu$, allows for a gauge invariant operator mixing the two field    
 strengths. 
In particular, the quadratic Lagrangian for the two gauge fields is given by
\bea
\label{eq:lag_mixing}
\mathcal L_\textrm{kin} = - \frac{1}{4} \hat F_{\mu\nu} \hat F^{\mu\nu} - \frac{1}{4} \hat F'_{\mu\nu} \hat F^{'\mu\nu} - \frac{\kappa}{2} \hat F'_{\mu\nu} \hat F^{\mu\nu},
\eea
with $\kappa$ being the kinetic mixing parameter. Since the parameterisation above may be inconvenient for practical computations, it is often useful to recast the kinetic Lagrangian into a diagonal form by removing the mixing operator through a rotation and rescaling of the Abelian fields. This transformation, while diagonalising Eq.~(\ref{eq:lag_mixing}), introduces a non-diagonal term in the interactions such that the covariant derivative may be written as
\bea
\mathcal D_{\mu} = \partial_\mu + \ldots + i g_1 Y B_\mu + i (\tilde g Y +  g' z) B'_\mu,
\eea
where $Y$ and $z$ are, respectively, the hypercharge and the $U(1)'$ charge, and $B_\mu,B'_\mu$ are the rotated fields. 
The parameter $\tilde g$ replaces $\kappa$ and describes the mixing between the two Abelian groups while $g'$ is the usual gauge coupling associated to the extra Abelian symmetry $U(1)'$.

Due to the mixing term in the gauge covariant derivative, after spontaneous symmetry breaking, the EW Vacuum Expectation Value 
(VEV) contributes to the $U(1)'$ breaking even if the Higgs sector is neutral under the new Abelian symmetry. For instance, in a scenario with only one Higgs doublet, the neutral gauge boson mass matrix can be extracted from the Higgs Lagrangian and reads as
\bea
- \mathcal L_\textrm{Higgs} = \frac{v^2}{8} (g_2 W_\mu^3 - g_1 B_\mu - g_\Phi B'_\mu)^2 + \frac{m_{B'}^2}{2} B_\mu^{'2} + \ldots,
\eea
where $g_\Phi = \tilde g + 2 z_\Phi g'$ with $z_\Phi$ being the $U(1)'$ charge of the SM Higgs or a combination of charges in multi-Higgs doublet scenarios. 
As stated above, a non-vanishing $g_\Phi$ can be achieved either by the non-zero $U(1)'$ charges of the Higgs sector, $z_\Phi \neq 0$, or by the presence of the kinetic mixing $\tilde g \neq 0$. Both of them contribute to a $Z - Z'$ mass mixing. 
The mass term $m_{B'}^2$ represents a possible source for the $Z'$ mass from a SM neutral sector. This can be realised, for instance, by the VEV $v'$ of a SM-singlet complex scalar $\chi$, with a $z_\chi$ charge under $U(1)'$. In this case $m_{B'} = g' z_\chi v'$. We remark here that, for our purposes, it is not necessary to specify the origin of the $B'$ mass term and other mechanisms, beside Spontaneous Symmetry Breaking  (SSB) with a complex scalar, can be also envisaged. 
Moreover, the mixing in the neutral gauge sector is only triggered by the $g_\Phi$ parameter and, as such, is unaffected by the details of the scalar sector in which the $B'$ mass term is generated. 

The diagonalisation of the mass matrix provides the relation between the interaction and the mass eigenstates and is described by the rotation matrix 
\bea
\left( \begin{array}{c} B^\mu \\ W_3^\mu \\ B'^\mu \end{array} \right) = \left( \begin{array}{ccc} 
\cos \theta_w & - \sin \theta_w \cos \theta' & \sin \theta_w \sin \theta' \\
\sin \theta_w & \cos \theta_w \cos \theta' & - \cos \theta_w \sin \theta' \\
0 & \sin \theta'  & \cos \theta'  
\end{array} \right)
\left( \begin{array}{c} A^\mu \\ Z^\mu \\ Z'^\mu \end{array} \right)
\eea
where $\theta_w$ is the usual weak mixing angle and $\theta'$ is a new mixing angle, with $-\pi/4 \le \theta' \le \pi/4$, defined as \cite{Accomando:2016sge}
\bea
\label{eq:mixing1Higgs}
\tan 2 \theta' = \frac{2  g_\Phi g_Z}{ g_{\Phi^2} + 4 m_{B'}^2/v^2 - g_Z^2},
\eea
where $g_Z = \sqrt{g_1^2 + g_2^2}$ is the EW coupling, $g_\Phi = \tilde g + 2 z_\Phi g'$ and $g_{\Phi^2} = g_{\Phi}^2$. The masses of the $Z$ and $Z'$ gauge bosons are then given by
\bea
\label{eq:ZZpmass1Higgs}
M_{Z,Z'} = g_Z \frac{v}{2} \left[ \frac{1}{2} \left( \frac{g_{\Phi^2} + 4 m_{B'}^2/v^2 }{g_Z^2} + 1\right) \mp \frac{g_{\Phi}}{\sin 2 \theta' \, g_Z} \right]^\frac{1}{2}.
\eea
For a light and weakly interacting $Z'$, namely $g', \tilde g \ll g_Z$ and $m_{B'}^2 \ll v^2$, the mixing angle and the masses can be expanded at leading order as
\bea
\label{eq:expansion}
M_Z^2 \simeq \frac{1}{4} g_Z^2 v^2 \,, \qquad M_{Z'}^2 \simeq m_{B'}^2\,, \qquad \tan 2 \theta' \simeq - 2 \frac{ g_\Phi}{g_Z}  \,.
\eea
While the SM $Z$ mass is correctly reproduced by the EW VEV, the mass of the $Z'$ is controlled by the $m_{B'}$ parameter or, equivalently, by the VEV $v'$ of the SM-singlet $\chi$ which is then given by $v' = M_{Z'}/(g' z_\chi)$. The $Z'$ massless limit for $m_{B'} = 0$ is naively expected since if SSB is turned off in the scalar sector, no scalar degrees of freedom can provide the longitudinal component of a massive $Z'$. For a 17 MeV $Z'$ with $g' \sim 10^{-3}$ the VEV of $\chi$ is $v' \sim 10$ GeV.  

The expansions in Eq.~(\ref{eq:expansion}) are applicable if the Higgs sector is populated by only one $SU(2)$ doublet, as in the SM. 
This assumption can be obviously relaxed and more Higgs doublets can be implemented. 
We show, indeed, in the following sections that this possibility leads to an interesting phenomenology in the $Z'$ sector and provides alternative solutions to the $\Be$ anomaly. 

For instance, in a scenario with two $SU(2)$ doublet scalars, $\Phi_1$ and $\Phi_2$ with the same hypercharge $Y=1/2$ and two different charges $z_{\Phi_1}$ and $z_{\Phi_2}$ under the extra $U(1)'$, the diagonalisation of the neutral gauge mass matrix is obtained through the mixing angle $\theta'$ in Eq.~(\ref{eq:mixing1Higgs}) with
\bea
g_\Phi &=& (\tilde g + 2 g' z_{\Phi_1}) \cos^2 \beta +  (\tilde g + 2 g' z_{\Phi_2})  \sin^2 \beta \,, \nn \\
g_{\Phi^2} &=& (\tilde g + 2 g' z_{\Phi_1})^2 \cos^2 \beta + (\tilde g + 2 g' z_{\Phi_2})^2 \sin^2 \beta \,.
\eea
The angle $\beta$ is defined as usual as $\tan \beta = v_2/v_1$ with $v^2 = v_1^2 + v_2^2$. 
In the small coupling limit the $Z'$ mass is given by
\bea
M_{Z'}^2 \simeq m_{B'}^2 + \frac{v^2}{4} {g'}^2 \left( z_{\Phi_1} - z_{\Phi_2} \right)^2 \sin^2(2 \beta),
\eea
which, differently from the previous case, is non-vanishing even when $m_{B'} \simeq 0$ due to mismatch between $z_{\Phi_1}$ and $z_{\Phi_2}$. 
In the limit in which there is no contribution from the dark scalar sector, one finds for $M_{Z'} \simeq 17$ MeV and $v \simeq 246$ GeV, $\tilde g \sim g' \sim 10^{-4}$. 
Interestingly, as we will show below, the same order of magnitude of the gauge couplings is required to explain the $\Be$ anomaly with a $Z'$ gauge boson characterised by axial-vector couplings. 

In summary, for the case of one Higgs doublet, we showed that the limit $m_{B'} \ll v$ leads to $M_{Z'} \simeq m_{B'}$ with the SM Higgs sector playing no role in the generation of the $Z'$ mass. In contrast, in a multi-Higgs scenario, like in a 2-Higgs Doublet Model (2HDM), if $z_{\Phi_1} \neq z_{\Phi_2}$, the symmetry breaking of the $U(1)'$ 
can actually be realised without any contribution from the dark scalar sector, namely with $v' = 0$. In fact, 
the longitudinal degree of freedom of the $Z'$ is provided by the typical CP-odd state of the 2HDM spectrum which, differently from standard constructions, 
is characterised by a missing pseudoscalar field among the physical states. Before moving to this 2HDM realisation, though, we ought to discuss the $Z'$ interactions with the SM fermions emerging from the present construct.

\subsection{The $Z'$ interactions with the SM fermions}
The interactions between the SM fermions and the $Z'$ gauge boson are described by the Lagrangian $\mathcal L_\textrm{int} = - J^\mu_{Z'} Z'_\mu$ where the gauge current is given by
\bea
J^\mu_{Z'} = \sum_f \bar \psi_f \gamma^\mu \left( C_{f, L} P_L + C_{f, R} P_R \right) \psi_f
\eea
with coefficients
\bea
C_{f,L} &=& - g_Z s' \left( T^3_f - s_w^2 Q_f \right) + (\tilde g Y_{f, L} + g' z_{f, L}) \, c' \,, \nn \\  
C_{f,R} &=& g_Z s_w^2 s' Q_f + (\tilde g Y_{f, R} + g' z_{f, R}) \, c' \,.
\eea
In the previous equations we have adopted the shorthand notation $s_w \equiv \sin \theta_w$, $c_w \equiv \cos \theta_w$, $s' \equiv \sin \theta'$ and $c' \equiv \cos \theta'$ and
introduced $Y_f$ the hypercharge, $z_f$ the $U(1)'$ charge, $T^3_f$ the third component of the weak isospin and $Q_f$ the electric charge.
Analogously, the vector and axial-vector components of the $Z'$ interactions are \cite{DelleRose:2017xil}
\bea
C_{f, V} &=& \frac{C_{f, R} + C_{f, L}}{2} = \frac{1}{2} \left[ 
- g_Z s' (T^3_f - 2 s_w^2 Q_f) + c' \tilde g (2 Q_f - T^3_f) + c' g' (z_{f, L} + z_{f,R})
\right] \,, \nn \\
C_{f, A} &=& \frac{C_{f, R} - C_{f, L}}{2} = \frac{1}{2} \left[ 
 (g_Z s'  + \tilde g c' ) T^3_f - c' g' (z_{f, L} - z_{f, R}) 
\right] \,,
\eea
The vector and axial-vector coefficients simplify considerably in the limit $g', \tilde g \ll g_Z$. By noticing that $s' \simeq - g_\Phi/g_Z$, we get
\bea
\label{eq:CVA_expanded}
C_{f, V} &\simeq&    \tilde g  c_w^2 \, Q_f + g'  \left[ z_\Phi (T^3_f - 2 s_w^2 Q_f)  + z_{f,V} \right] \,, \nn \\
C_{f, A} &\simeq&  g' \left[   -  z_\Phi \, T^3_f  +   z_{f,A} \right] \,,
\eea
where we have introduced the vector and axial-vector $U(1)'$ charges $z_{f,V/A} = 1/2(z_{f,R} \pm z_{f,L})$
and $z_\Phi$ can be either the $U(1)'$ charge of the Higgs or $z_{\Phi_1} \cos^2 \beta +  z_{\Phi_2} \sin^2 \beta$ in a 2HDM scenario. 

The $Z'$ couplings are characterised by the sum of three different contributions. 
The kinetic mixing $\tilde g$ induces a vector-like term proportional to the Electro-Magnetic (EM) current which is the only source of interactions when all the SM fields are neutral under $U(1)'$. In this  case the $Z'$ is commonly dubbed \emph{dark photon}. The second term is induced by the $z_\Phi$, the $U(1)'$ charge in the Higgs sector, and leads to a \emph{dark Z}, namely a gauge boson mixing with the SM $Z$ boson. Finally there is the standard gauge interaction proportional to the fermionic $U(1)'$ charges $z_{f,V/A}$. 

We can delineate different scenarios depending on the structure of the axial-vector couplings of the $Z'$ boson. 
In particular, the $C_{f,A}$ coefficients can be suppressed with respect to the vector-like counterparts (see also \cite{Kahn:2016vjr}). 
This is realised, for instance, when only one $SU(2)$ doublet is considered and the gauge invariance of the Yukawa Lagrangian under the new Abelian symmetry is enforced.
Indeed, the latter requires the $U(1)'$ charge of the Higgs field to satisfy the conditions
\bea
\label{eq:yukawa_gaugeinv}
z_\Phi = z_{Q} - z_{d} = - z_{Q} + z_{u} = z_L - z_e \,.
\eea
Inserting the previous relations into Eq.~(\ref{eq:CVA_expanded}), we find $C_{f, A} \simeq 0$ which describes a $Z'$ with only vector interactions with charged leptons and quarks.
We stress again that the suppression of the axial-vector coupling is only due to the structure of the scalar sector, which envisions only one $SU(2)$ doublet, and the gauge invariance of the Yukawa Lagrangian.  This feature is completely unrelated to the $U(1)'$ charge assignment of the fermions, the requirement of anomaly cancellation and the matter content potentially needed to account for it. 

In contrast, in the scenario characterised by two Higgs doublets, the axial-vector couplings of the $Z'$ are, in general, of the same order of magnitude of the vector ones and the cancellation between the two terms of $C_{f, A}$ in Eq.~(\ref{eq:CVA_expanded}) is not achieved regardless of the details of the Yukawa Lagrangian (such as which type 2HDM). 
The same result can be achieved if a single Higgs doublet is considered but the conditions in Eq.~(\ref{eq:yukawa_gaugeinv}) are not satisfied as in scenarios in which the fermion masses are generated radiatively or through horizontal symmetries.

To summarise, we can identify three different situations that can provide a light $Z'$ with interactions potentially explaining the Beryllium anomaly. In all of them, the SM is extended by an additional Abelian gauge group.

\noindent
$1.$ The SM scalar sector is unchanged, being characterised by only one Higgs doublet. In this case the mass of the $Z'$ is entirely generated in the dark sector. The Yukawa Lagrangian preserves the SM structure and its gauge invariance under the $U(1)'$ necessary implies that the $Z'$ has only vector interactions with the SM fermions at leading order in the couplings $\tilde g, g'$.

\noindent
$2.$ The SM scalar sector is extended by an additional Higgs doublet. Even though the Yukawa Lagrangian is invariant under the local $U(1)'$ symmetry, the cancellation between the two terms in $C_{f,A}$ in Eq.~(\ref{eq:CVA_expanded}) does not occur and both the vector and axial-vector couplings of the $Z'$ are non-vanishing. The mass of the $Z'$ acquires contribution from both the dark and the EW sectors.

\noindent
$3.$ The SM scalar sector is characterised by a single Higgs doublet but the constraints in Eq.~(\ref{eq:yukawa_gaugeinv}) are avoided by relying on more complicated Yukawa structures. As such, the cancellation providing $C_{f,A} \simeq 0$ is not realised and the vector and axial-vector interactions of the $Z'$ are of the same order of magnitude.

\noindent
We will discuss the three scenarios in the following sections focusing on their implications in the interpretation of the $\Be$ anomaly.

Before concluding this section we briefly go through the conditions required by the cancellation of gauge and gravitational anomalies which strongly constrain the charge assignment of the SM spectrum under the extra $U(1)'$ gauge symmetry. These conditions can be eventually combined with the requirement of  gauge invariance of the Lagrangian responsible for the generation of the fermion masses which may also involve non-renormalisable operators. 
We will also allow for extra SM-singlet fermions which can be easily interpreted as right-handed neutrinos. 
We assign the charges $z_Q$ and $z_L$ for the $SU(2)$ quark and lepton doublets, $z_u, z_d, z_e$ for the corresponding right-handed components and $z_{\nu}$ for the $n_R$ right-handed neutrinos. We obtain the following gauge and gravitational anomaly cancellation conditions:
	\begin{align}
	\label{eq:anomaly}
	& U(1)'SU(3)SU(3): & \sum_i^{3} (2 z_{Q_i} - z_{u_i} - z_{d_i}) = 0 \,,  \nn \\
	& U(1)'SU(3)SU(3): &  \sum_i^{3} \, ( 3 z_{Q_i} +  z_{L_i})  = 0 \,, \nn \\
	& U(1)'U(1)_YU(1)_Y: & \sum_i^{3} \left( \frac{z_{Q_i}}{6} - \frac{4}{3} z_{u_i} - \frac{z_{d_i}}{3}  +   \frac{z_{{L_i}}}{2} - z_{e_i}\right) = 0 \,, \nn \\
	& U(1)'U(1)'U(1)_Y: & \sum_i^{3} \left( z_{Q_i}^2 - 2 z_{u_i}^2 + z_{d_i}^2      - z_{{L_i}}^2 + z_{e_i}^2 \right) = 0 \,, \nn \\
	& U(1)'U(1)'U(1)': & \sum_i^{3} \left( 6 z_{Q_i}^3 - 3 z_{u_i}^3 - 3 z_{d_i}^3 +  2 z_{{L_i}}^3 - z_{e_i}^3 \right)  + \sum_i^{n_R} z_{\nu _i}    = 0 \,, \nn \\
	& U(1)'GG: & \sum_i^{3} \left( 6 z_{Q_i} - 3 z_{u_i} - 3 z_{d_i}  + 2 z_{{L_i}} - z_{e_i} \right) + \sum_i^{n_R} z_{\nu _i}    = 0 .
	\end{align}
A simple solution is found for instance in the family universal case with $n_R = 3$ and $z_{\nu_{i}} = z_{\nu}$ and it is defined in terms of only two $U(1)'$ charges, $z_Q$ and $z_u$ as shown in Tab. \ref{tab:charges}.
As an example, the $U(1)_{B-L}$ is reproduced by $z_Q = z_u = 1/3$ while the sequential $U(1)'$ is obtained for $z_Q = 1/6$ and $z_u = 2/3$.
\begin{table}
\centering
\begin{tabular}{|c|c|c|c|c|}
\hline
		& $SU(3)$ & $SU(2)$ & $U(1)_Y$ & $U(1)'$ \\ \hline
$Q_L$	&  3		&	2	& 1/6		&	$z_Q$ \\
$u_R$	&  3		&	1	& 2/3		&	$z_u$ \\
$d_R$	&  3		&	1	& -1/3	&	$2 z_Q - z_u$\\
$L$		&  1		&	2	& -1/2	&	$-3 z_Q$ \\
$e_R$	&  1		&	1	& -1		&	$-2 z_Q - z_u$ \\
$\nu_R$	&  1		&	1	& 0		&	$- 4 z_Q + z_u$ \\
\hline
\end{tabular}
\caption{Family universal charge assignment in the $U(1)'$ extension of the SM. \label{tab:charges}}
\end{table}

\subsection{$Z'$ with vector couplings}
\label{sec:ZpVector}
The simplest $U(1)'$ extension of the SM, which may account for an extra neutral light gauge boson potentially explaining the $\Be$ anomaly, is characterised by a single Higgs doublet. 
As already explained above, the gauge invariance of the Yukawa interactions fixes the $U(1)'$ charge of the Higgs to satisfy the restrictions in Eq.~(\ref{eq:yukawa_gaugeinv})
thus leading to a suppression of the $Z'$ axial-vector couplings to the quarks and charged leptons with respect to the vector ones. \\
In this scenario, the anomalous internal pair creation transition of the excited stated of the Beryllium described by the normalised BR is given by
\bea
\frac{{\rm BR}(^8{{\rm Be}^*} \to X + {^8{\rm Be}})}{{\rm BR}(^8{{\rm Be}^*} \to \gamma + {^8{\rm Be}})}  = \frac{1}{e^2}(C_{p, V} + C_{n, V})^2 \frac{|\vec{k}_{Z'}|^3}{|\vec{k}_\gamma|^3}
\eea
in which any dependence from the nuclear matrix elements factors out in the ratio of BRs. Moreover, the partial decay width of the $Z'$ into SM fermions is
\bea
\label{eq:Zpdecaywidth}
\Gamma(Z' \rightarrow f \bar f) = \frac{M_{Z'}}{12 \pi} \sqrt{1 -  \frac{4 m_f^2}{M_{Z'}^2}} \left[ C_{f,V}^2 + C_{f,A}^2 + 2 ( C_{f,V}^2 - 2 C_{f,A}^2 ) \frac{m_f^2}{M_{Z'}^2} \right] \,.
\eea
Since $M_{Z'} \simeq 17$ MeV, the light $Z'$ can only decay into electrons and active neutrinos (assuming heavier right-handed neutrinos, if any). \\
While $C_{f, A} \simeq 0$, the explicit expressions of the vector couplings of the $Z'$ are
\bea
C_{p, V} &=& \tilde g c_w^2 - 2 g' z_H s_w^2 + g' (z_H + 3 z_Q) \,, \nn \\
C_{n, V} &=& - g' \left( z_H - 3 z_Q \right) \,, \nn \\
C_{e, V} &=& - \tilde g c_w^2 + 2 g' z_H s_w^2 - g' (z_H -  z_L)\,, \nn \\
C_{\nu, V} &=& - C_{\nu, A} = \frac{g'}{2} (z_H + z_L) \,,
\eea
where we have introduced the proton and neutron couplings $C_{p,V} = 2 C_{u,V} + C_{d,V}$, $C_{n,V} = C_{u,V} + 2 C_{d,V}$ and we have exploited the gauge invariance of the Yukawa Lagrangian.
Moreover, the cancellation of the anomaly in the $U(1)'SU(2)SU(2)$ triangle diagram given in Eq.~(\ref{eq:anomaly}) leads to $3 z_Q + z_L = 0$, namely $C_{\nu,V} = - 2 C_{n,V}$.  \\
The acceptable range of couplings is \cite{Feng:2016jff,Feng:2016ysn}
\bea
|C_{p, V}| &\lesssim& 1.2 \times 10^{-3} \, e \,, \nn \\
|C_{n, V}| &=& (2 - 10) \times 10^{-3} \, e \,, \nn \\
|C_{e, V}| &=& (0.2 - 1.4) \times 10^{-3} \, e \,, \nn \\
\sqrt{|C_{\nu, V} C_{e, V}|} &\lesssim& 3 \times 10^{-4} e\,,
\eea
where $\textrm{BR}(Z' \to e^+ e^-) = 1$ has been assumed. The first two conditions ensure that the Atomki anomaly is correctly reproduced while avoiding, at the same time, the strong constraint from the $\pi^0 \to Z' \gamma$ decay. As the coupling to proton is smaller than the corresponding one to neutron, the $Z'$ realising this particular configuration has been dubbed \emph{protophobic}. The bound on the electron coupling is mainly obtained from KLOE2, $(g-2)_e$ and beam dump experiments, while the neutrino coupling is constrained by neutrino scattering off electrons
at the Taiwan EXperiment On Neutrinos (TEXONO) \cite{Deniz:2009mu}. Reinterpreting the bounds obtained in \cite{Bilmis:2015lja}, where a $B-L$ scenario without mixing has been considered, for a general vector-like $Z'$, one can show that the $C_\nu$ coupling must be much smaller than the typical value of $C_{n,V}$ required to explain the $\Be$ anomaly, thus invalidating the $C_{\nu,V} = - 2 C_{n,V}$ condition required by the consistency of this simple model. 

A possible way to suppress the neutrino coupling, without affecting the neutron one, could be to invoke the presence of additional neutral fermionic degrees of freedom, charged under the $U(1)'$ symmetry and mixed to the left-handed neutrinos, so that the effective coupling of the $Z'$ to the physical neutrino mass eigenstate would be significantly reduced. 
This mixing is commonly realised in the seesaw mechanism, which is naturally envisaged in the Abelian extension considered here since right-handed neutrinos are required to cancel the gauge anomalies, but it can hardly account for the bounds determined by the neutrino-electron scattering experiments. 
Such a strategy has been discussed in \cite{Feng:2016ysn}, however, here we show two alternative solutions based on the exploitation of the $Z'$ axial-vector interactions.

\section{Explanation of the Beryllium anomaly with a family universal $U(1)'$}

In this section we investigate the explanation of the Atomki anomaly in a scenario characterised by an extra $U(1)'$ model and two Higgs doublets. 

One possibility studied as a solution to the Atomki anomaly is a well-known realisation of the scalar potential and Yukawa interactions with two scalar doublets is the so-called type-II in which the up-type quarks couple to one Higgs  (conventionally chosen to be $\Phi_2$) while the down-type quarks couple to the other ($\Phi_1$).  The constraint from anomaly cancellation arising from the $U(1)'SU(3)SU(3)$ triangle diagram together with the gauge invariance of the Yukawa Lagrangian require $2 z_Q - z_d - z_u = z_{\Phi_1} - z_{\Phi_2} = 0$, in the type-II scenario. In order to satisfy this condition with $z_{\Phi_1} \neq z_{\Phi_2}$, extra coloured states are necessarily required which will bring additional terms into the anomaly cancellation conditions and so the equation above will be modified and no longer require equal Higgs charges under the new gauge group. These  states must be vector-like under the SM gauge group and chiral under the extra $U(1)'$. This option has been explored in detail in \cite{Kahn:2016vjr}. In this work, the constraints on new vector bosons with axial vector couplings in a family-universal scenario which includes extra coloured states to cancel anomaly conditions is considered. In this review focus on a different, more minimal scenario than this, which does not require additional states, but modifies the scalar theory to affect the condition of anomaly cancellation.

The gauge invariance condition above is modified when the scalar sector reproduces the structure of the type-I 2HDM in which only one ($\Phi_2$) of the two Higgs doublets participates in the Yukawa interactions. In this theory, the corresponding Lagrangian is the same as the SM one and its gauge invariance simply requires $z_{\Phi_2} = - z_Q + z_u = z_{Q} - z_d = z_{L} - z_e$, without constraining the $U(1)'$ charge of $\Phi_1$, in the type-I scenario. In this way, we allow for gauge invariance even when $z_{\Phi _1} \neq z_{\Phi _2}$. Differently from the type-II scenario in which extra coloured states are required to build an anomaly-free model, in the type-I case the UV consistency of the theory can be easily satisfied introducing only SM-singlet fermions as demanded by the anomaly cancellation conditions of the $U(1)'U(1)'U(1)'$ and $U(1)'GG$ correlators. Nevertheless, the mismatch between $z_\Phi$ and $z_{f,A}=\pm z_{\Phi_2}/2$ (for up-type and down-type quarks, respectively) prevents $C_{f,A}$ to be suppressed 
and the $Z'$ interactions are given by \cite{DelleRose:2017xil},
\begin{align}
&C_{u, V} =  \frac{2}{3} \tilde g  c_w^2  + g'  \left[ z_\Phi \left(\frac{1}{2} -  \frac{4}{3}  s_w^2 \right)  + z_{u,V} \right], \nn\\
&C_{u, A} = - \frac{g'}{2} \cos^2 \beta (z_{\Phi_1} - z_{\Phi_2})  \,, \nn \\
&C_{d, V} =  -\frac{1}{3} \tilde g  c_w^2  + g'  \left[ z_\Phi \left(-\frac{1}{2} +  \frac{2}{3} s_w^2  \right)  + z_{d,V} \right],  \nn\\
&C_{d, A} = \frac{g'}{2} \cos^2 \beta  (z_{\Phi_1} - z_{\Phi_2}) \,, \nn \\
&C_{e, V} =  - \tilde g  c_w^2  + g'  \left[ z_\Phi \left(-\frac{1}{2} +  2 s_w^2  \right)  + z_{e,V} \right],  \nn\\
&C_{e, A} =  \frac{g'}{2} \cos^2 \beta  (z_{\Phi_1} - z_{\Phi_2}) \,, \nn \\
&C_{\nu, V} = - C_{\nu, A} = \frac{g'}{2} \left( z_{\Phi} + z_L \right) .
\label{couplings}
\end{align}

As pointed out in \cite{Feng:2016ysn}, the contribution of the axial-vector couplings to the $\Be^* \rightarrow \Be \, Z'$ decay is proportional to $|\vec{k}_{Z'}|/M_{Z'} \ll 1$, where $\vec{k}_{Z'}$ is the momentum of the $Z'$, while the vector component is suppressed by $|\vec{k}_{Z'}|^3/M_{Z'}^3$. Therefore, in our case, being $C_{f,V} \sim C_{f,A}$, we can neglect the effects of the vector couplings of the $Z'$ and their interference with the axial counterparts. 
For a $Z'$ with only axial-vector couplings to quarks, the transition $\Be^* \rightarrow \Be \, Z'$ is described by the partial width \cite{Kozaczuk:2016nma}
\bea
\Gamma = \frac{k}{18 \pi} \left( 2 + \frac{E_k^2}{M_{Z'}^2} \right) \left| a_n \langle 0 || \sigma^n || 1 \rangle + a_p \langle 0 || \sigma^p || 1 \rangle \right|^2,
\eea
where the neutron and proton coefficients $a_n = (a_0-a_1)/2$ and $a_p = (a_0+a_1)/2$ are defined in terms of 
\bea
a_0 &=& \left( C_{u,A} + C_{d,A}\right) \left( \Delta u^{(p)}  + \Delta d^{(p)} \right) + 2 \, C_{s,A} \Delta s^{(p)} \,, \nn \\
a_1 &=& \left( C_{u,A} - C_{d,A}\right) \left( \Delta u^{(p)}  - \Delta d^{(p)} \right)\,,
\eea
with $\Delta u^{(p)} = 0.897(27)$, $\Delta d^{(p)} = -0.367(27)$ and $\Delta s^{(p)} = - 0.026(4)$ \cite{Bishara:2016hek}. 
The reduced nuclear matrix elements of the spin operators have been computed in \cite{Kozaczuk:2016nma} and are given by $\langle 0 || \sigma^n || 1 \rangle = -0.132 \pm 0.033$, $\langle 0 || \sigma^p || 1 \rangle = -0.047 \pm 0.029$ for the isoscalar $\Be^* \rightarrow \Be$ transition and $\langle 0 || \sigma^n || 1 \rangle = -0.073 \pm 0.029$, $\langle 0 || \sigma^p || 1 \rangle = 0.102 \pm 0.028$ for the isovector $\Be^{*'} \rightarrow \Be$ transition.

\begin{figure}
\centering
\includegraphics[width=9.cm,height=9.cm]{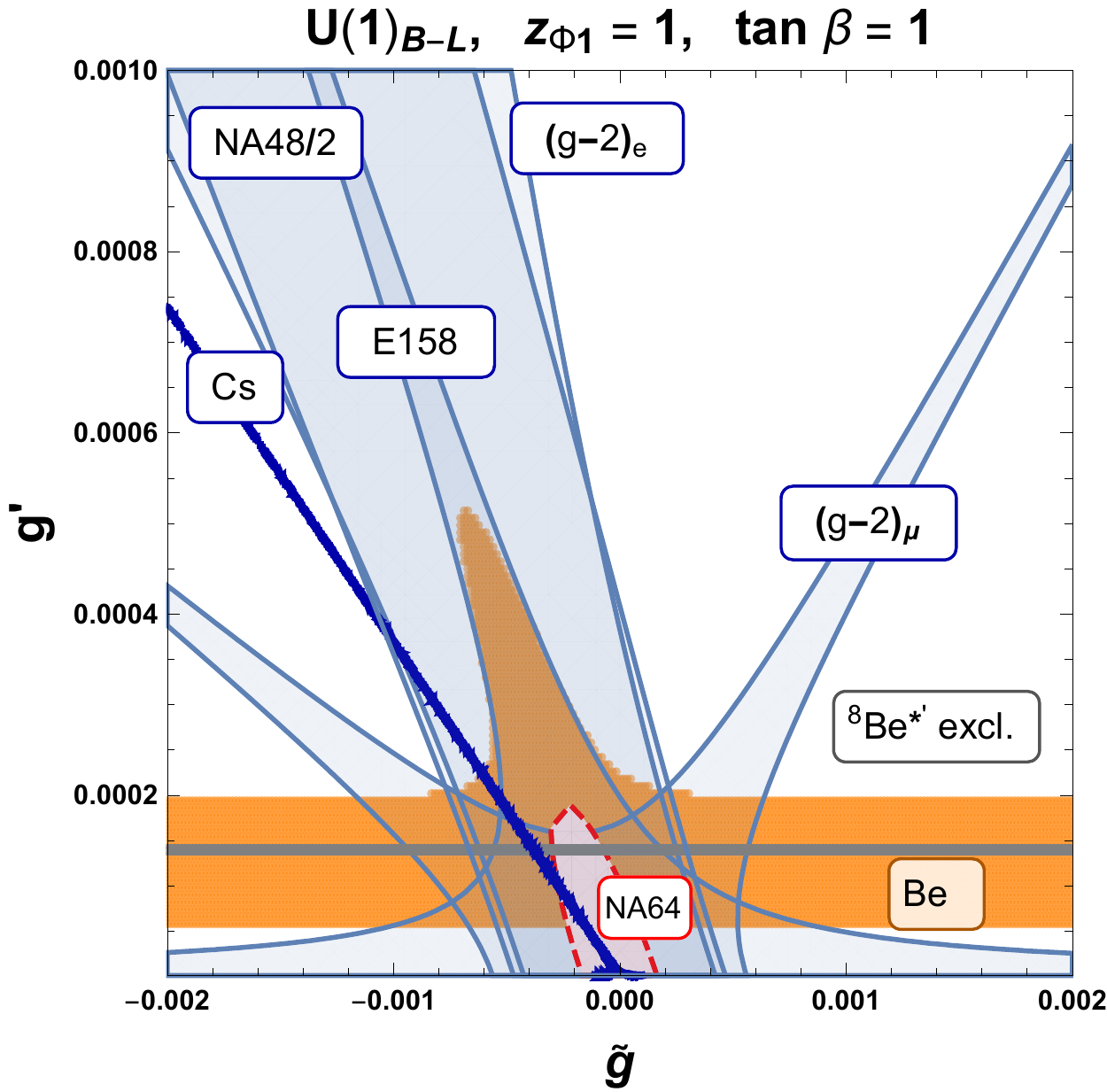}
\caption{
Allowed parameter space (orange region) explaining the anomalous $\Be^*$ decay. 
The white region above is excluded by the non-observation of the same anomaly in the ${\Be^{*}}'$ transition. Also shown (shaded regions) is the allowed parameter space by the $g-2$ of electrons and muons and the M{\o}ller scattering at SLAC E158 and pion decay from NA48/2. The beam dump experiment NA64 allows parameter space outside the red shaded region with dashed line. Finally, the blue line selects values of $g'$ and $\tilde g$ compatible with the weak nuclear charge measurement of Cesium. The horizontal grey band delineates values of $g'$ for which the $Z'$ mass is solely generated by the SM vev.
\label{fig:typeI}}
\end{figure}

Notice that the axial couplings of the quarks and, therefore, the width of the $\Be^* \rightarrow \Be \, Z'$ decay are solely controlled by the product $g' \cos^2 \beta$ while the kinetic mixing $\tilde g$ only affects the $\textrm{BR}(Z' \rightarrow e^+e^-)$ since the $Z' \rightarrow \nu \nu$ decay modes are allowed (we assume that the $Z' \rightarrow \nu_R \nu_R$ decays are kinematically closed). 
For definiteness, we consider a $U(1)_\textrm{B-L}$ charge assignment with $z_{Q_{L}} =z_{u_R}=1/3$, with other charges defined using Tab. \ref{tab:charges}, and $z_{\Phi_2} = 0$, $z_{\Phi_1} = 1$ and $\tan \beta = 1$. 
Analogue results may be obtained for different $U(1)'$ charge assignments and values of $\tan \beta$.
We show in Fig.~\ref{fig:typeI} the parameter space explaining the Atomki anomaly together with the most constraining experimental results.

The orange region, where the $Z'$ gauge couplings comply with the best-fit of the $\Be^*$ decay rate in the mass range $M_{Z'} = 16.7 \, {\rm MeV} - 17.6 \, {\rm MeV}$ \cite{Krasznahorkay:2015iga,Feng:2016ysn}, encompasses the uncertainties on the computation of the nuclear matrix elements \cite{Kozaczuk:2016nma}. The 
region above it is excluded by the non-observation of the same transition in the isovector excitation ${\Be^{*}}'$ \cite{Krasznahorkay:2015iga}.
The horizontal grey band selects the values of $g'$ accounting for the $Z'$ mass in the negligible $m_{B'}$ case in which the $U(1)'$ symmetry breaking is driven by the two Higgs doublets.
Furthermore, among all other experimental constraints involving a light $Z'$ that may be relevant for this analysis we have shown the most restrictive ones.

The strongest bound comes from the atomic parity violation in Cs and it represents a constraint on the product of $C_{e,A}$ and a combination of $C_{u,V}$ and $C_{d,V}$.
This bound can be avoided if the $Z'$ has either only vector or axial-vector couplings but in the general scenario considered here, it imposes severe constraints on the gauge couplings $g',\tilde g$ thus introducing a fine-tuning in the two gauge parameters.
We finally comment on the constraints imposed by neutrino-electron scattering processes \cite{Vilain:1994qy,Deniz:2009mu,Bellini:2011rx}, the strongest one being from $\bar \nu_e e$ scattering at the TEXONO experiment \cite{Deniz:2009mu}, which affect a combination of $C_{e, V/A}$ and $C_{\nu,V}$. 
As discussed above, in the protophobic scenario, in which the $Z'$ has only vector interactions, the constrained $\nu$ coupling to the $Z'$ boson is in high tension with the measured $\Be^*$ decay rate since $C_{\nu,V} = -2 C_{n,V}$ and a mechanism to suppress the neutrino coupling must be envisaged \cite{Feng:2016ysn}. 
This bound is, in general, alleviated if one attempts to explain the Atomki anomaly with a $Z'$ boson with axial-vector interactions since the required gauge couplings $g',\tilde g$ are smaller than the ones needed in the protophobic case. 
{
Neutrino couplings are also constrained by meson decays, like, for instance $K^\pm \rightarrow \pi^\pm \nu \nu$ which has been studied in \cite{Davoudiasl:2014kua} and where it has been shown that the corresponding constraint is relaxed by a destructive interference effect induced by the charged Higgs. As the results presented in \cite{Davoudiasl:2014kua} relies on the Goldstone boson equivalence approximation,
we have computed the full one-loop corrections to the $K^\pm \rightarrow \pi^\pm Z'$ process in the $U(1)'$-2HDM scenario. 
The results are in agreement with the estimates in \cite{Davoudiasl:2014kua}.
In our setup, for $g' \sim 10^{-4}$ and $\tan \beta = 1$, $M_{H^\pm} \sim 600$ GeV can account for the destructive interference quoted above between the $W^\pm$ and $H^\pm$ loops. 
For instance, we find $\textrm{BR}(K^\pm \rightarrow \pi^\pm Z' \rightarrow \pi^\pm \nu \nu) \simeq 0.1 \, \textrm{BR}(K^\pm \rightarrow  \pi^\pm \nu \nu)_\textrm{exp}$ for $M_{H^\pm} \sim 615$ GeV with $\textrm{BR}(Z' \rightarrow \nu\nu) \simeq 30\%$ which is the maximum value for the invisible $Z'$ decay rate in the allowed region (orange and grey shaded area) shown in Fig.~\ref{fig:typeI}. 
A similar constraint arises from the $B$ meson decay to invisible but is less severe than the one discussed above \cite{Patrignani:2016xqp}.
The $B^\pm \rightarrow K^\pm Z'$ process is characterised by the same loop corrections appearing in $K^\pm \rightarrow \pi^\pm Z'$, with the main difference being the dependence on the Cabibbo-Kobayashi-Maskawa (CKM) matrix elements. Therefore, the suppression effect induced by the charge Higgs mass affects both processes in the same region of the parameter space, thus ensuring that the bound from the invisible $B$ decays is satisfied once the constraint from the analogous $K$ meson decay is taken into account.
} \\

\section{Explanation of the Beryllium anomaly with a family non-universal $U(1)'$}
The final alternative is to consider a single Higgs doublet, as with the SM, but non-standard Yukawa interactions, to allow axial couplings through the violation of Eq.~(\ref{eq:yukawa_gaugeinv}), as done in \cite{DelleRose:2018eic}. This is done for the first two generations of fermions and the third has SM-like gauge-invariance, motivated by $\mathcal{O}(1)$ couplings. We begin by modifying the Yukawa couplings for the first two generations as follows,
\bea
- \mathcal{L}_{Yuk} &=& \Gamma^{u} \dfrac{\chi^{n_{ij}}}{M^{n_{ij}}} \overline{Q}_{L,i}\tilde{H}u_{R,j} 
+ \Gamma^{d} \dfrac{\chi^{l_{ij}}}{M^{l_{ij}}} \overline{Q}_{L,i} H d_{R,j} \nonumber \\
&+&\Gamma^{e} \dfrac{\chi^{m_{ij}}}{M^{m_{ij}}} \overline{L}_{i} H e_{R,j}+ h.c.,
\eea
where the exponent, $n_{ij}$, of the non-renormalisable scale, $M$, is defined by the $U(1)'$ charges of the fields, such that these new Yukawa terms are gauge invariant. Subsequently, one may obtain fermion masses either at tree-level or radiatively by the method of Ref. 
\cite{Froggatt:1978nt}\footnote{Lagrangians of this form have been used to motivate solutions to the flavour problem, so it may be of interest to investigate whether this $U(1)'$ may explain the allowed masses and mixings, but we perform no such careful investigation here.}.
There are several models which motivate radiative mass generation for the lighter generations, as done in \cite{Demir:2005ti}, alternatively, there exist mass generation dynamics by horizontal symmetries, as in \cite{Froggatt:1978nt}. We do not specify these dynamics, and simply leave an effective approach. We finally enforce that the first two generations are flavour universal, differing from the third, $z_{i_1}=z_{i_2}$ for $i=\{Q,u_R,d_R,L,e_R\}$, where the condition (\ref{eq:yukawa_gaugeinv}) is not applied to $z_{i_{1,2}}$. We now consider further constraints on the charge assignment. 
We also enforce the chiral anomaly cancellation conditions in Eq.~(\ref{eq:anomaly}), which will be satisfied by solely the fermionic content of the SM, supplemented by two right-handed neutrinos.

Our remaining constraints on the charge assignment are motivated by the non-observation of BSM physics. As discussed above, there are strong constraints on coupling to neutrinos, which would  enhance processes such as $K^\pm \rightarrow \pi ^\pm \nu \nu$ \cite{Davoudiasl:2014kua}, as well as electron-neutrino interactions, measured by the TEXONO experiment \cite{Feng:2016ysn,Deniz:2009mu,Bilmis:2015lja,Khan:2016uon}. To avoid these stringent constraints, we therefore impose no couplings to the neutrinos, i.e., $C_{V,\nu} = C_{A,\nu} = 0$. This subsequently yields a relation between the neutrino and Higgs charges,
\begin{equation}
	z_{L_1}=z_{L_2}=z_{L_3}=-z_{H}.
\end{equation}
Another constraint is to require that one indeed does have axial couplings for the up/down quarks to the $Z'$, as required to explain the anomaly,
\begin{align}
	&-z_{Q_{1,2}} -z_H +z_{u_{1,2}} \neq 0, \\
	&-z_{Q_{1,2}} + z_H +z_{d_{1,2}} \neq 0.
\end{align}
Our final constraint is from the atomic parity violation in Cs. As can be seen from other solutions, this provides a stringent bound on models with axial couplings for electrons. We thus also forbid interactions of this kind, and due to requiring universality for the first two generations, this will also forbid axial couplings for the muon,
\begin{align}
C_{e,A} = C_{\mu,A} = 0. 
\end{align}
Preventing the appearance of these axial couplings will also help to avoid bounds from both $(g-2)_e$ and avoid worsening the discrepancy in $(g-2)_\mu$.

Combining all these constraints yields a single, unique charge assignment. We have a normalisation choice, and choose to set $z_H =1$. This unique choice is shown in Tab. \ref{tab:charges1HDM}.

\begin{table}[!t]
	\centering
	\begin{tabular}{|c|c|c|c|c|}
		\hline
		& \multirow{2}{*}{$SU(3)$} & \multirow{2}{*}{$SU(2)$} & \multirow{2}{*}{$U(1)_Y$} & \multirow{2}{*}{$U(1)'$} \\
		&&&&\\ \hline \vspace{-1em}
		&&&&\\
		$Q_{1}$	&  3		&	2	& 1/6	&	$1/3$  \\
		$Q_{2}$	&  3		&	2	& 1/6	&	$1/3$  \\
		$Q_{{3}}$	&  3		&	2	& 1/6	& $1/3$ \\
		$u_{R_{1}}$	&  3		&	1	& 2/3	&	$-2/3$ \\
		$u_{R_{2}}$	&  3		&	1	& 2/3	&	$-2/3$ \\
		$u_{R_{3}}$	&  3		&	1	& 2/3	&	$4/3$  \\
		
		$d_{R_{1}}$	&  3		&	1	& -1/3	&	$4/3$ \\
		$d_{R_{2}}$	&  3		&	1	& -1/3	&	$4/3$\\
		$d_{R_{3}}$	&  3		&	1	& -1/3	&	$-2/3$ \\
		$L_{1}$		&  1		&	2	& -1/2	&	$-1$ \\
		$L_{2}$		&  1		&	2	& -1/2	&	$-1$  \\
		$L_{3}$		&  1		&	2	& -1/2	&	$-1$ \\
		$e_{R_{1}}$	&  1		&	1	& -1		&	$0$ \\
		$e_{R_{2}}$	&  1		&	1	& -1		&	$0$ \\
		$e_{R_{3}}$	&  1		&	1	& -1		&	$-2$ \\
		$H$	&  1		&	2	& 	1/2	&	$1$  \\
		\hline
	\end{tabular}
	\caption{Charge assignment of the SM particles under the family-dependent (non-universal) $U(1)'$. This numerical charge assignment  satisfies the discussed anomaly cancellation conditions, enforces a gauge invariant Yukawa sector of the third generation and family universality in the first two fermion generations as well as  no coupling of the $Z'$ to the all neutrino generations.}
	\label{tab:charges1HDM}
\end{table}

Now, we consider constraints on the new gauge coupling, and gauge-kinetic mixing parameters $(g', \tilde{g})$, given this charge selection. Unlike the previous scenarios considered, since this is family non-universal, one finds tree level FCNCs, which should be analysed. In diagonalising the quarks into the mass basis, off-diagonal couplings are generated, due to different coupling strengths between the first two and third quark generations. We now discuss the consequences of this on experimental observables. We begin with $K \rightarrow \pi e^+ e^-$ through a tree-level exchange of an on-shell $Z'$. There are no contributions to the $\mu ^+ \mu ^-$ decay as $M_{Z'} \sim 17$ MeV $< 2m_{\mu}$. There are stringent limits from LHCb \cite{Aaij:2015dea}, though these are inapplicable in our case due to the small invariant mass of the $e^+ e^-$ pair. There is only sensitivity to energies above 20 MeV, due to photon conversion in the detector, and so energy resolution strongly degrades around these invariant masses. It is possible that future upgrades will lower this threshold and thus act as a discovery tool, or to disprove this scenario.

Another flavour observable is from meson mixing measurements. We begin with $B^0 - \bar{B}^0$, following the procedure as done in \cite{Becirevic:2016zri}, but now assuming a much lighter propagator than their scenario, $P \equiv (m_{B} ^2 - M_{Z'}^2)^{-1} \simeq m_{B}^{-2}$, as opposed to their $P \simeq M_{Z'}^{-2}$. One subsequently finds the requirement 
\begin{equation}
	|g^{L(R)} _{sb}| \lesssim 10^{-6},
\end{equation}
where (assuming Minimal Flavour Violation (MFV) in the quark sector and using CKM matrix elements),
\begin{align}
	g^L _{sb} &= g' \Big( V_{\rm CKM} ^T ~\textrm{Diag}(z_{Q_1}  , z_{Q_1}  , z_{Q_3} )~ V_{\rm CKM} \Big) _{23} ,\\
	g^R _{sb} &= g' \Big( V_{\rm CKM} ^T ~\textrm{Diag}(z_{u_{R_1}}  , z_{u_{R_1}}  , z_{u_{R_3}} )~ V_{\rm CKM} \Big) _{23} ,
\end{align} 
Since our charge assignment is family universal for LH quarks, $g^L _{sb}=0$, see Tab. \ref{tab:charges1HDM}, only the right-handed sector will contribute to the FCNC. This is suppressed by CKM factors, $g^R _{sb} \propto V_{tb} V_{ts}$, and so one finds a condition on the couplings, $g', \tilde g \lesssim 10^{-4}$. \\
Proceeding in a similar faction but for $K - \bar{K}$ oscillations will yield a weaker constraint on the couplings. Although the propagator suppression is less severe, $P \simeq m_K ^{-2} > m_{B} ^{-2}$ , the CKM suppression is much stronger, $g_{sd} ^R \propto V_{td} V_{ts}$, and one finds the constraint $g', \tilde g \lesssim 10^{-3}$. In this review, we do not perform a full flavour analysis, but require these approximate constraints.

\begin{figure}[h]
	\centering
	\includegraphics[width=0.7\linewidth]{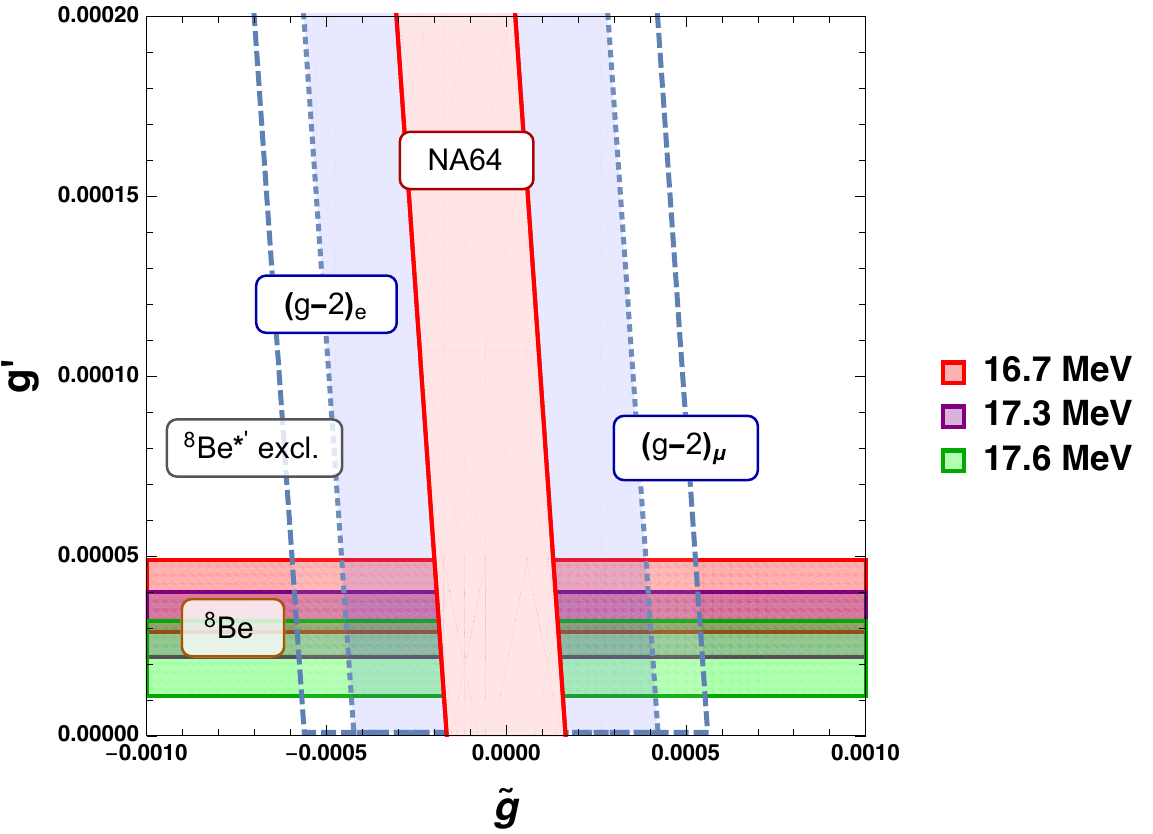}
	\caption{Allowed parameter space mapped on the $(g',\tilde{g})$ plane 
		explaining the anomalous ${}^8 \textrm{Be}^*$ decay for $Z'$ solutions with mass 16.7 (red), 17.3 (purple) and 17.6 (green) MeV. The white regions are excluded by the non-observation of the same anomaly in the ${}^8 \textrm{Be}^{*'}$ transition. Also shown are the constraints from $(g-2)_\mu$, to be within the two dashed lines; $(g-2)_e$, to be inside the two dotted lines (shaded in blue) and the electron beam dump experiment, NA64, to be outside the shaded red region, which lies between the two solid lines. The surviving parameter space lies at small positive and negative $\tilde{g}$ (though not at $\tilde{g}=0$), inside the shaded blue region which overlaps the Atomki anomaly solutions.}
	\label{fig:Final_Region_NA64}
\end{figure}

Finally, we present the allowed parameter space in Fig. \ref{fig:Final_Region_NA64} for this scenario with one Higgs doublet extended by a $U(1)'$, with a charge assignment shown in Tab. \ref{tab:charges1HDM}. The red, purple and green bands show regions which can explain the Atomki anomaly for 16.7, 17.3 and 17.6 MeV $Z'$ masses, respectively. These overlap in places and are independent of $\tilde{g}$ as the axial coupling depends solely on $g'$ and BR$(Z' \rightarrow e^+ e^-)=1$ everywhere. These bands have upper bounds due to the non-observation of the ${}^8 $Be$^{*'}$ anomaly. Also shown on the plot are the bounds from $(g-2)_\mu$, where the allowed region is inside the dashed line and $(g-2)_e$, where the allowed region is shaded in blue inside the dotted lines. In addition, the allowed region from NA64 is also shown, where one should be outside the red shaded region. The overall allowed region is therefore between the NA64 and $(g-2)_e$ lines, in the overlap shaded in blue. The other experimental constraints (electron positron collider (KLOE2), Moller scattering (E158), pion decay (NA48/2), E141, and atomic parity violation of Cs), similar to $(g-2)_\mu$, do not limit the allowed parameter space in blue, and are not shown on the plot.

\begin{table}[!t]
	\centering
	\begin{tabular}{c | c}
		$M_{Z'} ~(\textrm{MeV}) $ & ${\mathcal B}$ \\ \hline \vspace{-1em}
		&\\ 
		16.7 & $5.8 \times 10^{-6}$ \\
		17.3 & $2.3 \times 10^{-6}$ \\
		17.6 & $5.0 \times 10^{-7}$
	\end{tabular}
	\caption{Solutions to the Atomki anomaly, with best fit mass value (16.7 MeV) from \cite{Krasznahorkay:2015iga} and subsequent alternative masses (17.3 MeV and 17.6 MeV) from \cite{Feng:2016ysn} along with the corresponding ratio of BRs, ${\mathcal B}$, as defined in Eq. (\ref{eq:BeAnomaly}).}
	\label{tab:Br}
\end{table}

\begin{figure}[h]
	\centering
	\includegraphics[width=0.6\linewidth]{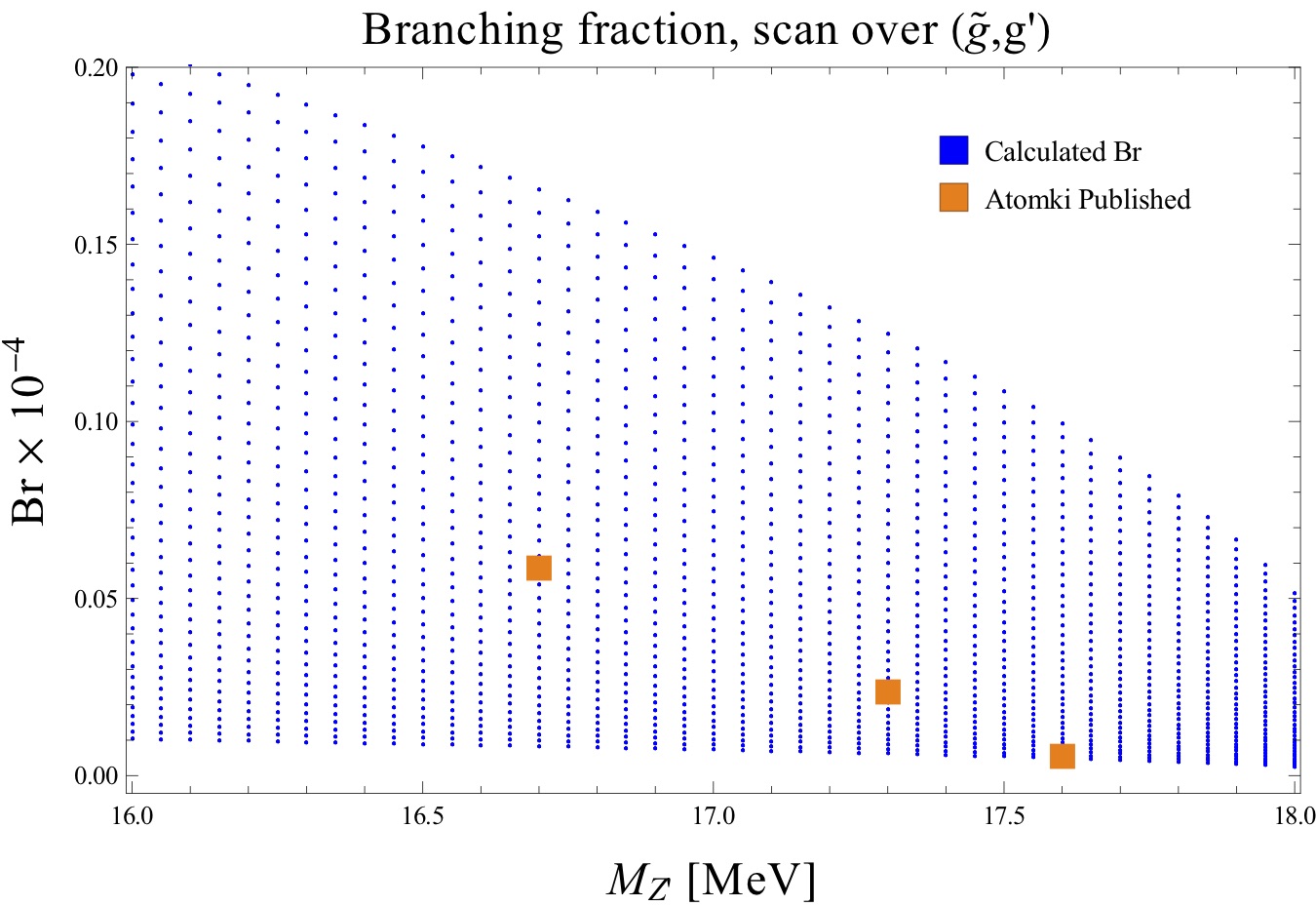}
	\caption{Values of the upper limit ${\mathcal B}$ (lower limits are smaller than the scale of the plot), as defined in Eq. (\ref{eq:BeAnomaly}), versus the mass of the $Z'$ obtained scanning over the allowed parameter space in $(g',\tilde{g})$, obtained from Fig.~\ref{fig:Final_Region_NA64} for each mass step taken (in blue). The Atomki collaboration solutions are also shown (in orange).}
	\label{fig:BRPlotDensity_high}
\end{figure}

Fig.~\ref{fig:BRPlotDensity_high} shows the quantity $\mathcal{B}$, as defined in Eq.~(\ref{eq:BeAnomaly}), over a range of $Z'$ masses. For each fixed mass value, a scan is performed over $(g',\tilde{g})$, in a range compatible with other experimental constraints, and the Atomki anomaly (i.e., over the dark blue and coloured regions in Fig. \ref{fig:Final_Region_NA64}). For each scanned point in $\{ M_{Z'},~g'.~\tilde{g} \}$, there is a range of branching ratios, due to uncertainties in the Nuclear Matrix Elements (NMEs). This lower limit for all points is lower than the Atomki branching ratios, so only the upper ${\mathcal B}$ is of importance, and this is plotted. Also drawn, in orange, is the required branching ratio, as published by the Atomki collaboration, see Tab. \ref{tab:Br}. A given point is then allowed if the upper ${\mathcal B}$ limit lies above the orange dots. For larger $M_{Z'}$ values, the largest ${\mathcal B}$ decreases, and a larger number of the scanned points lie above the Atomki points. This suggests that at higher masses, there is slightly more parameter space available for the 17.6 MeV solution, in comparison to the 16.7 MeV one. This is reflected in the slightly different widths shown in Fig. \ref{fig:Final_Region_NA64}.

\section{Conclusions}

While there remains the possibility that the Atomki anomaly can be explained as a statistical fluctuation combined with yet unknown nuclear physics properties and/or unforeseen experimental conditions, the fact that presently such an effect has been determined with a $6.8\sigma$ significance, including a near-perfect fit of both the mass and angular excesses to the possibility of a new particle with a mass of about 17 MeV  been produced, calls for a thorough investigation of plausible theoretical explanations. 

With this in mind, in this review, we have presented particle
physics scenarios that extend the SM to include the presence of either a spin-0 (pseudoscalar, $A$) boson or a spin-1 (axial-vector, $Z'$) boson, both of which can be made compliant with a variety of experimental data. Assuming the standard Lagrangian structures describing $A$ and $Z'$ interactions with SM fermionic currents in both the lepton and quark sectors, we have  determined the required couplings of such bosons to explain the Beryllium data.

As for the theoretically embeddings of these solutions, we can conclude the following. A light pseudoscalar state can appear in models with extended Higgs sectors 
in which an approximate ungauged global symmetry is spontaneously
broken, examples of which include (type-II) 2HDMs with a SM-singlet near the Peccei-Quinn
or $R$-symmetric limit, although in this case isospin breaking effects and non-universality in  the Yukawa couplings of the new state to electrons and $d$-quarks  must be allowed for.  As for light gauge bosons with significant axial-vector couplings, two possible theoretical frameworks have been proven to be viable. Both require an additional $U(1)'$ group mixing with the SM one, $U(1)_Y$. In one case, which retains the SM Higgs sector, a family non-universal set of $Z'$ couplings to the known fermions must be invoked. In the other case, $Z'$ couplings to quarks and fermions of the SM can be retained in their universal form, yet this requires an enlarged Higgs sector, which we have identified as possibly being a type-I 2HDM. Needless to say, these two theoretical frameworks were constructed in presence of gauge invariance and anomaly cancellations plus they do not require isospin breaking.

While the above list of possible theoretical setups is clearly not exhaustive, it at least provides somewhat minimal frameworks (only containing enlarged Higgs and gauge sectors, possibly including heavy neutrinos but no exotic particles) within which further data upcoming from the Atomki experiment can be interpreted to pave the way for  more dedicated phenomenological studies, which may in turn lead to  refinements on the theoretical side.     

\section*{Acknowledgements}
The work of LDR and SM is supported in part by the NExT Institute. SM also acknowledges partial financial contributions from the STFC Consolidated Grant ST/L000296/1. Furthermore, the work of LDR has been supported by the STFC/COFUND Rutherford International Fellowship Programme (RIFP). SJDK and SK have received support under the H2020-MSCA grant agreements InvisiblesPlus (RISE) No. 690575 and Elusives (ITN) No. 674896. In addition SK was partially supported by the STDF project 13858. All authors acknowledge support under the H2020-MSCA grant  agreement NonMinimalHiggs (RISE)
No. 645722.


\end{document}